\documentclass[aps, prb, reprint, superscriptaddress, floatfix]{revtex4-2}
\usepackage{graphicx}
\usepackage{dcolumn}
\usepackage{bm}
\usepackage{physics}
\usepackage{amsfonts}
\usepackage{mathrsfs}
\usepackage{subfigure}
\usepackage[svgnames]{xcolor}
\usepackage[colorlinks=true,linkcolor=NavyBlue,anchorcolor=red,citecolor=NavyBlue, urlcolor=NavyBlue]{hyperref}
\usepackage{color}
\usepackage{xcolor}
\DeclareMathAlphabet{\pazocal}{OMS}{zplm}{m}{n}
\usepackage{amsmath}
\usepackage{amssymb}
\usepackage[T1]{fontenc}

\usepackage{color}
\usepackage{makecell}

\begin{document}

\preprint{APS/123-QED}

\title{Machine Learning Modeling of Charge-Density-Wave Recovery After Laser Melting}

\author{Sankha Subhra Bakshi}
\affiliation{Department of Physics, University of Virginia, Charlottesville, Virginia, 22904, USA}

\author {Yunhao Fan}
\affiliation{Department of Physics, University of Virginia, Charlottesville, Virginia, 22904, USA}

\author {Gia-Wei Chern}
\affiliation{Department of Physics, University of Virginia, Charlottesville, Virginia, 22904, USA}

\begin{abstract}
We investigate the nonequilibrium dynamics of a laser-pumped two-dimensional spinless Holstein model within a semiclassical framework, focusing on the melting and recovery of long-range charge-density-wave order. Accurately describing this process requires fully nonadiabatic electron--lattice dynamics, which is computationally demanding due to the need to resolve fast electronic motion over long time scales. By analyzing the structure of the lattice force during nonequilibrium evolution, we show that the force naturally separates into a smooth quasi-adiabatic component and a residual bath-like contribution associated with fast electronic fluctuations. The quasi-adiabatic component depends only on the instantaneous local lattice configuration and can be efficiently learned using machine-learning techniques, while a minimal Langevin description of the bath term captures the essential features of the recovery dynamics. Combining these elements enables efficient and scalable simulations of long-time nonequilibrium dynamics on large lattices, providing a practical route to access driven correlated systems beyond the reach of direct nonadiabatic approaches.
\end{abstract}

\date{\today}

\maketitle

\section{Introduction}
\label{sec:intro}

Understanding the nonequilibrium dynamics of strongly coupled electron–lattice systems remains a central challenge in condensed matter physics. Ultrafast pump–probe experiments have revealed that ordered phases such as charge-density waves (CDWs) can be selectively melted by intense laser excitation~\cite{OS1,OS2,OS3,OS4,OS5} and subsequently recover over time scales spanning many orders of magnitude~\cite{PICO1,PICO2,PICO3,PICO4,PICO5,PICO6}. These experiments expose rich dynamical pathways—including nonthermal melting, transient symmetry restoration, and slow reformation of long-range order—that are fundamentally inaccessible in equilibrium. From a theoretical perspective, capturing such phenomena requires a framework that treats electronic and lattice degrees of freedom on equal footing and goes beyond adiabatic or quasistatic approximations.

A paradigmatic platform for exploring this physics is the Holstein model~\cite{Holstein1959,Holst2,Holst3}, in which local lattice distortions couple directly to the electronic density and stabilize CDW order at half filling. In equilibrium, or in near-adiabatic regimes, lattice dynamics can be efficiently described by forces derived from the instantaneous electronic ground state~\cite{Sclass_holst1,Sclass_holst2,Sclass_holst3,Sclass_holst4,Sclass_holst5}. Crucially, these forces are local functionals of the lattice configuration, a property that has recently enabled the development of machine-learning (ML) force-field approaches that bypass repeated diagonalization of large electronic Hamiltonians~\cite{cheng23a,ML2,ML3,ML4}. Such methods dramatically reduce computational cost while retaining near–first-principles accuracy, opening the door to large-scale simulations of electron–phonon systems.

Out of equilibrium, however, this simplification breaks down. Following a strong laser excitation, the electronic system is driven far from equilibrium, and the force acting on the lattice depends on the full time-dependent electronic density matrix rather than solely on the instantaneous lattice configuration~\cite{HolstPhoto1,HolstPhoto2,HolstPhoto3,HolstPhoto4,HolstPhoto5,HolstPhoto6}. This force generically encodes long-ranged and history-dependent electronic correlations. Accurately capturing the resulting nonadiabatic dynamics requires explicit time evolution of the electronic degrees of freedom on fast microscopic time scales over very long durations, leading to severe numerical bottlenecks. As a consequence, direct simulations are typically restricted to small system sizes or short times, limiting theoretical access to the long-time recovery dynamics of broken-symmetry phases.

In this work, we develop a linear-scaling machine-learning (ML) force-field framework that enables efficient simulations of charge-density-wave (CDW) recovery dynamics in laser-driven electron–lattice systems. Focusing on the nonequilibrium Holstein model, we show that, despite the fully nonadiabatic electronic evolution induced by strong laser excitation, the force acting on the lattice admits a physically transparent decomposition. Specifically, it separates into a smooth quasi-adiabatic component that depends primarily on the instantaneous lattice configuration and a residual bath-like contribution arising from nonequilibrium electronic fluctuations. Crucially, we find that the quasi-adiabatic force remains a local and well-behaved functional of the lattice degrees of freedom even far from equilibrium, making it amenable to data-driven modeling. We demonstrate that this dominant force component can be learned accurately and efficiently using ML force-field models trained directly on nonequilibrium trajectories.

Our approach is motivated by, and builds upon, the rapidly growing literature on machine-learning force fields in atomistic simulations. ML force-field models have achieved near–quantum-mechanical accuracy at dramatically reduced computational cost by learning high-dimensional potential energy surfaces from density functional theory or other electronic-structure methods~\cite{behler07,bartok10,li15,shapeev16,behler16,botu17,smith17,zhang18,deringer19,mcgibbon17,suwa19,chmiela17,chmiela18,sauceda20,unke21}. By bypassing explicit electronic-structure calculations during dynamical simulations, these approaches have extended molecular dynamics to system sizes and time scales far beyond those accessible to direct ab initio methods~\cite{Marx2009}. However, the vast majority of existing ML force-field frameworks are formulated for equilibrium or near-equilibrium conditions, where forces derive from a well-defined potential energy surface associated with the electronic ground state.

Here, we extend the ML force-field paradigm beyond this equilibrium setting by adopting a graph neural network (GNN) architecture tailored to nonequilibrium electron–lattice dynamics. Rather than learning forces from an underlying equilibrium potential, our approach targets the quasi-adiabatic component of the force extracted from laser-driven, time-dependent electronic evolution. The GNN framework provides a natural representation of lattice systems in terms of local environments and interactions defined on a graph, enabling computational cost that scales linearly with system size. Moreover, discrete lattice translation symmetry and point-group symmetries can be incorporated explicitly through symmetry-aware graph construction and equivariant message-passing rules. This yields a compact and physically constrained representation of the quasi-adiabatic force field, ensuring transferability across system sizes and lattice configurations. As a result, our approach makes it possible to access long-time recovery dynamics of laser-melted CDW phases. 


The remainder of the paper is organized as follows. In Sec.~II we introduce the model
and semiclassical framework for nonequilibrium dynamics and discuss a typical melting and recovery dynamics. In Sec.~III we analyze the
structure of the nonadiabatic force and its decomposition into adiabatic and bath-like
components. In Sec.~IV we demonstrate how this insight enables an efficient
machine-learning description of the lattice force, In Sec.~V we apply it to benchmark the long-time CDW
recovery. We conclude in Sec.~VI with a discussion of implications and
outlook.

\section{Non-adiabatic dynamics for the Holstein model}
\label{sec:model}
We study the spinless Holstein model at half filling on a two-dimensional square lattice,
\begin{equation}
\hat{\mathcal{H}}=\hat{\mathcal{H}}_{\mathrm{e}}+\hat{\mathcal{H}}_{\mathrm{L}}+\hat{\mathcal{H}}_{\mathrm{eL}} ,
\end{equation}
where
\begin{equation}
\hat{\mathcal{H}}_{\mathrm{e}}=-t_{\rm nn}\sum_{\langle ij\rangle}
\left(\hat{c}^{\dagger}_{i}\hat{c}_{j}+\text{H.c.}\right)
\end{equation}
describes nearest-neighbor hopping of spinless fermions,
\begin{equation}
\hat{\mathcal{H}}_{\mathrm{L}}=\sum_i\left(\frac{\hat{P}_i^2}{2m}
+\frac{1}{2}m\Omega^2\hat{Q}_i^2\right)
\end{equation}
is the local phonon Hamiltonian, and
\begin{equation}
\hat{\mathcal{H}}_{\mathrm{eL}}=-g\sum_i \hat{n}_i\hat{Q}_i
\end{equation}
captures the electron–lattice coupling. Here $\hat{n}_i=\hat{c}^\dagger_i\hat{c}_i$,
$\hat{Q}_i$ and $\hat{P}_i$ denote the lattice displacement and its conjugate momentum,
$m$ is the phonon mass, and $\Omega$ the bare phonon frequency. The coupling is written
in particle–hole symmetric form, ensuring $\langle \hat{n}_i\rangle=\tfrac{1}{2}$ and
$\langle \hat{Q}_i\rangle=Q_0={g}/{2m\Omega^2}$ in equilibrium at half filling.

Motivated by earlier work demonstrating that the semiclassical Holstein model captures
the essential physics of charge-density-wave order~\cite{Holst3,Johnston2013,Weber2018,Nowadnick12,Costa2020, golez12}, we adopt a semiclassical description
to study post-quench dynamics in the collisionless regime. We approximate the full
time-dependent state by a product ansatz,
\begin{equation}
\ket{\Gamma(t)}=\ket{\Phi(t)}\otimes\ket{\Psi(t)},
\end{equation}
with $\ket{\Phi(t)}$ and $\ket{\Psi(t)}$ denoting the phonon and electronic many-body
states, respectively.

Treating the phonons at the mean-field level, we further assume
$\ket{\Phi(t)}=\prod_i\ket{\phi_i(t)}$. The lattice degrees of freedom are then fully
characterized by the classical variables
$Q_i=\langle\hat{Q}_i\rangle$ and $P_i=\langle\hat{P}_i\rangle$.
Their dynamics follows from the Ehrenfest equations evaluated in the product state,
leading to
\begin{equation}
    \label{eq:EOM_Q}
\frac{dQ_i}{dt}=\frac{P_i}{m}, \qquad
\frac{dP_i}{dt}=g \langle \hat{n}_i\rangle-m\Omega^2 Q_i .
\end{equation}
These are Newtonian equations for the classical lattice fields, coupled self-consistently
to the evolving electronic density.

Within the semiclassical framework, the electronic sector evolves in the time-dependent
lattice background $\{Q_i(t)\}$. We describe the electronic state in terms of the single-particle
density matrix
\begin{equation}
\rho_{ij}(t)=\langle \hat{c}^\dagger_j \hat{c}_i \rangle ,
\end{equation}
whose equation of motion follows from the Heisenberg equation,
\begin{equation}
\frac{d\rho}{dt}= i \left[\rho, h(\{Q_i\})\right].
\end{equation}
Here the effective single-particle Hamiltonian is
\begin{equation}
h_{ij} = -t_{ij} + g Q_i \delta_{ij},
\end{equation}
with $t_{ij}=t_{\rm nn}$ for nearest neighbors and zero otherwise. The coupled evolution of
$\rho_{ij}(t)$ and $Q_i(t)$ constitutes a fully nonadiabatic dynamics of electrons and lattice
degrees of freedom.

At half filling, the model exhibits a charge-density-wave (CDW) instability at wave vector
$\mathbf{K}=(\pi,\pi)$, characterized by the staggered lattice and charge order parameter
$Q_{\pi,\pi}(t)$. We perturb this ordered state using an external laser pulse, modeled as a
classical electric field
\begin{equation}
\mathbf E(t)= \hat{\mathbf e}\, \mathcal{A}
\exp\!\left[-\frac{(t-t_0)^2}{2\sigma_p^2}\right]
\sin(\Omega_p t),
\end{equation}
with central frequency $\Omega_p$, temporal width $\sigma_p$, polarization $\hat{\mathbf e}$ peak amplitude $\mathcal{A}$,
and centered at time $t_0$.

\begin{figure*}[!t]
    \centering
    \includegraphics[width=0.8\linewidth]{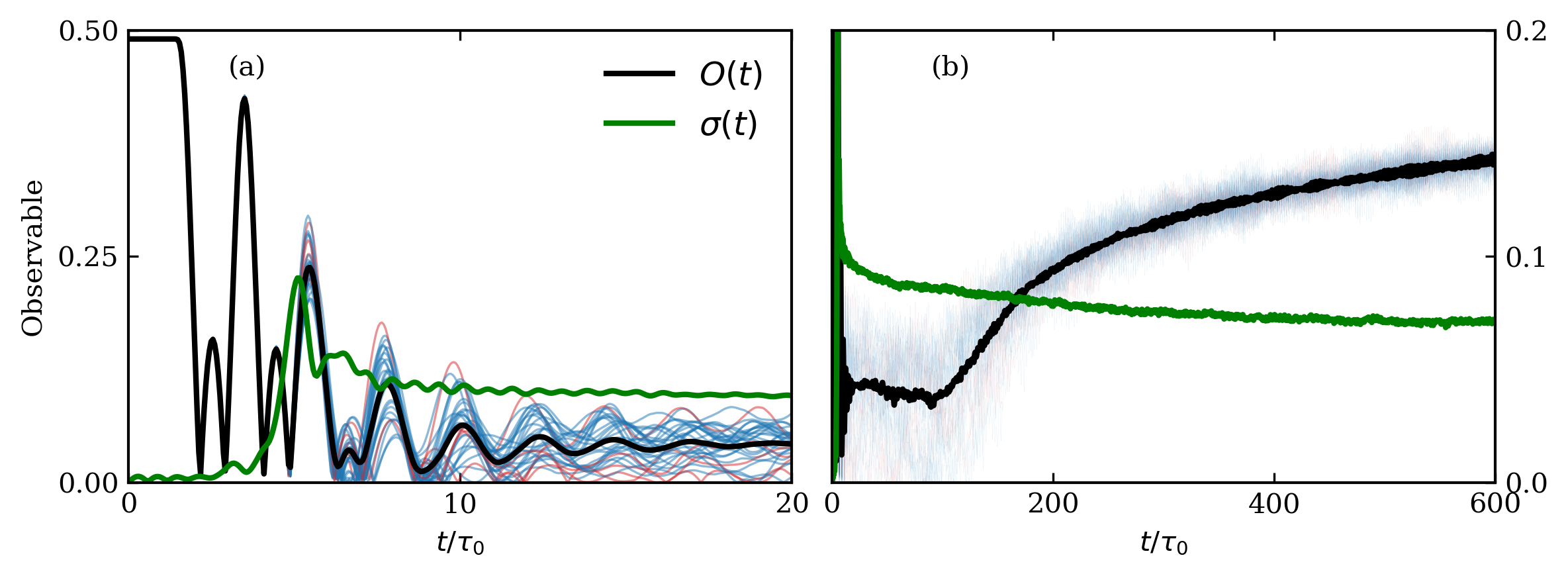}
   \caption{%
Pump-induced melting and recovery of the lattice order parameter.
Colored curves show individual stochastic realizations of order parameter,
with blue (red) indicating positive (negative) long-time values.
The black solid line denotes the ensemble-averaged order parameter
$O(t) $, while the green solid line shows the ensemble-averaged spatial fluctuation
$\sigma(t)$.
The left panel shows early-time dynamics ($t/\tau_0 \leq 20$), highlighting the transient suppression of charge order after excitation, and the right panel displays the long-time recovery dynamics.
}
    \label{fig:orderparam}
\end{figure*}

The field couples to the electrons through the Peierls substitution in the hopping term,
\begin{equation}
t_{ij} \;\rightarrow\; t_{ij} \,
\exp\!\left[-i\int_{\mathbf r_i}^{\mathbf r_j} \mathbf A(t)\cdot d\mathbf r\right],
\end{equation}
where the vector potential $\mathbf A(t)$ is related to the electric field via
$\mathbf E(t)=-\partial_t \mathbf A(t)$.
A pulse with frequency and bandwidth exceeding the CDW gap can induce excitations across
the gap, leading to partial or complete melting of charge order.

Physically, the laser pulse creates electronic excitations that reduce the charge modulation,
to which the lattice responds dynamically. The outcome depends sensitively on the pulse
strength: weak excitation suppresses but does not destroy CDW order, strong excitation
leads to complete melting, while intermediate excitation produces loss of long-range order
followed by a slow recovery at long times.

We illustrate one such intermediate case in Fig.~\ref{fig:orderparam}. Throughout, we set
$t_{\rm nn}=-1$ and $g=\sqrt{3}$, corresponding to a dimensionless coupling
\begin{equation}
\lambda=\frac{g^2/(m\Omega^2)}{W}=0.375,
\end{equation}
where the bandwidth is $W=8|t_{\rm nn}|$ and $m\Omega^2=1$. In equilibrium, this yields a
CDW density modulation $\Delta n\simeq0.57$ (with $\Delta n=1$ for perfect charge order).

The dynamics involves two intrinsic time scales: the electronic time scale
$\tau_{\rm el}=1/|t_{\rm nn}|=1$ and the bare phonon time scale $\tau_0=1/\Omega$.
We work in the adiabatic regime characterized by the ratio
$r=\tau_0/\tau_{\rm el}=0.2$.

The laser parameters are chosen as $\mathcal{A}=0.6$, $\Omega_p=1.5|t_{\rm nn}|$,
$\sigma_p=\tau_0/10$, with polarization along the $\hat{e}=(x,y)$ direction.
All results are obtained on a square lattice of linear size $L=20$.

To seed inhomogeneity and allow domain formation, we introduce weak stochastic noise
in the initial lattice configuration $Q_i(t=0)$, corresponding to an effective temperature
$T\sim10^{-5}|t_{\rm nn}|$. Each stochastic realization is labeled by an index $\alpha$,
and the subsequent time evolution is denoted $Q_i^{\alpha}(t)$.

Figure~\ref{fig:orderparam} shows the resulting nonadiabatic dynamics of the global lattice
order parameter, obtained from the Fourier component of the lattice distortion field at
$\mathbf{K}=(\pi,\pi)$. For each stochastic realization $\alpha$, we compute
$Q^{\alpha}_{\pi,\pi}(t)$ and define the ensemble-averaged order parameter as
\begin{equation}
O(t)=\frac{1}{N_{\alpha}}\sum_{\alpha}\bigl|Q^{\alpha}_{\pi,\pi}(t)\bigr|.
\end{equation}
Representative trajectories of $\bigl|Q^{\alpha}_{\pi,\pi}(t)\bigr|$ are shown together with
the ensemble-averaged order parameter $O(t)$ (black solid line). The green solid line shows
the ensemble-averaged spatial fluctuation $\sigma(t)$.
Individual trajectories are colored in blue or red according to the sign of the long-time value of $Q^{\alpha}_{\pi,\pi}(t)$, reflecting the two $\mathbb{Z}_2$-related CDW states.

Following the laser excitation, the CDW order rapidly melts and
is effectively lost by $t\sim15\,\tau_0$. During this interval, the finite value of
$\sigma(t)$ indicates the presence of short-range charge correlations despite the absence
of global order.

\begin{figure*}[!ht]
    \centering
    \includegraphics[width=0.8\linewidth]{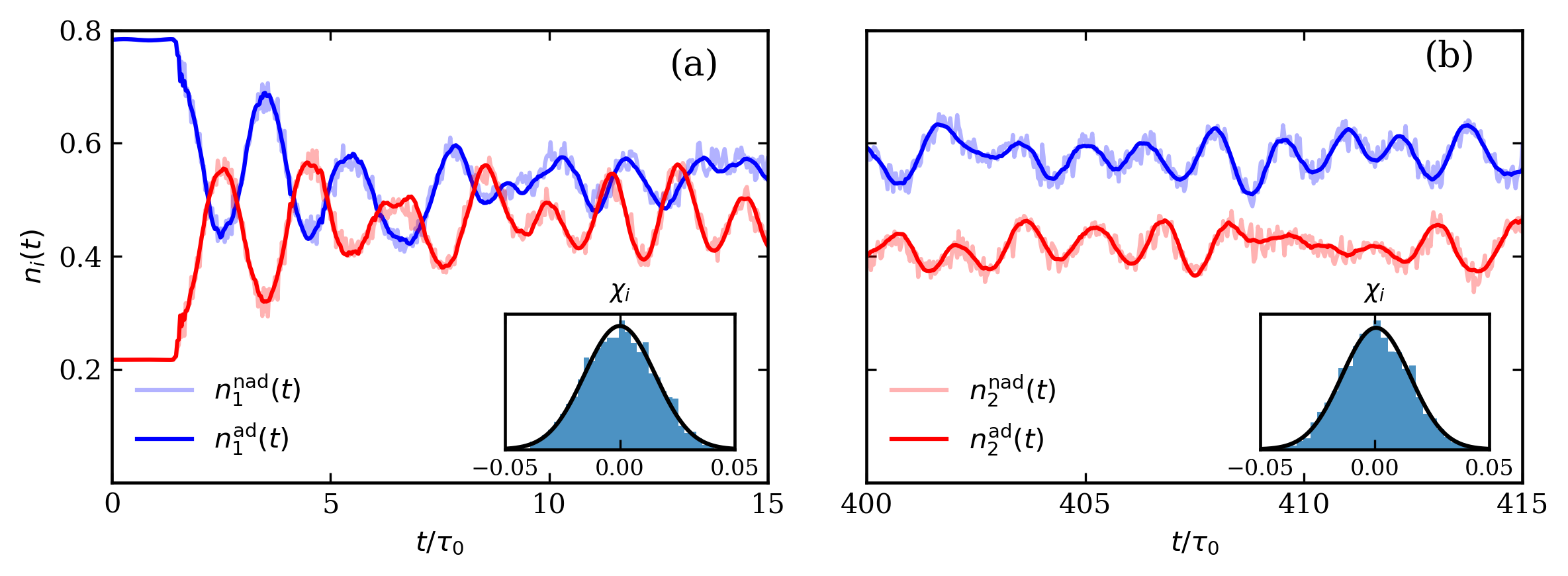}
    \caption{%
    Time evolution of the local density $n_i(t)$ for two representative sites.
    (a) Early-time dynamics ($t/\tau_0 \leq 15$), showing the comparison
    between the non-adiabatic evolution $n_i^{\mathrm{nad}}(t)$ (faint blue and red colors)
    and the adiabatic reference $n_i^{\mathrm{ad}}(t)$ (solid red and blue).
    (b) Late-time dynamics ($400 \leq t/\tau_0 \leq 415$).
    Insets in both panels show the distribution of
    $\chi_i = n_i^{\mathrm{ad}}(t) - n_i^{\mathrm{nad}}(t)$ within the corresponding
    time windows.
    }
    \label{fig:decomp}
\end{figure*}

At longer times, the system enters a recovery regime. Up to $t\sim100\,\tau_0$ the global
order parameter remains strongly suppressed, while the spatial fluctuations decay slowly.
Beyond this time scale, $O(t)$ begins to increase, signaling the gradual re-emergence
of long-range CDW order.

\section{Decomposition of the Nonequilibrium Force}
\label{sec:decomp}

The lattice dynamics is governed by the force acting on each phonon coordinate, as defined in Eq.~(\ref{eq:EOM_Q}), which consists of elastic and electronic contributions. The elastic component is purely classical and is obtained directly from the lattice Hamiltonian $\hat{\mathcal{H}}_L$,
\begin{eqnarray}
    F^{\rm elastic}_i = -m \Omega^2 Q_i,
\end{eqnarray}
and can be evaluated straightforwardly. Our primary interest lies in the electronic force, 
\begin{equation}
F^{\rm elec}_i(t) =  g \langle \hat{n}_i(t)\rangle .
\end{equation}
which couples the lattice to the electronic degrees of freedom. In equilibrium, evaluating this force requires computing the electronic density
$\langle \hat{n}_i \rangle$ in the instantaneous lattice background $\{Q_i\}$.
This involves diagonalizing the single-particle Hamiltonian of dimension
$N\times N$, with $N=L^2$, which constitutes the dominant numerical cost in the
force evaluation and quickly becomes a practical bottleneck as the system size
or the number of force evaluations increases.

In that case, a central observation is that the force $F_i$ is a local quantity: for a given site
$i$, it is primarily determined by the lattice configuration $\{Q_j\}$ in the
vicinity of $i$. This locality suggests that the force functional
$F_i[\{Q_j\}]$ can be represented using only local lattice information.
From a practical perspective, this opens the possibility of replacing repeated
global electronic structure calculations with a local surrogate description of
the force, trained to reproduce the exact result within the relevant region of
configuration space~\cite{cheng23a,ML2}.

Out of equilibrium, the numerical challenge takes a different form. Rather than
diagonalizing the Hamiltonian at each time step, one must explicitly evolve the
electronic density matrix $\rho_{ij}(t)$, which involves operations on objects of
size $N\times N$. While this avoids repeated diagonalizations, the dynamics
introduces a severe multiscale problem. In the adiabatic regime, the phonon time
scale $\tau_0 \gg \tau_{\rm el}$, and the recovery dynamics of interest typically
extends over hundreds of phonon periods.

As a result, accurate nonadiabatic simulations require evolving $\rho_{ij}(t)$
using very small time steps over extremely long durations. The total numerical
effort therefore grows rapidly with both system size and simulation time.
Unlike equilibrium settings, this bottleneck cannot be alleviated by exploiting
phonon slowness directly, since the electrons and lattice remain dynamically
coupled throughout the evolution.

An additional difficulty in the nonequilibrium dynamics arises from the
evaluation of the electronic contribution to the lattice force. 
\begin{align}
F_i^{\mathrm{nad}}(t) = g n_i^{\mathrm{nad}}(t) = g \rho_{ii}(t).
\end{align}
Here, we explicitly denote this force as $F^{\mathrm{nad}}$ to distinguish it from the quasi-adiabatic electronic force introduced below.
In contrast to equilibrium, the nonadiabatic density $n_i^{\mathrm{nad}}(t)$ does
not depend solely on the local lattice environment near site $i$, but is
determined by the full nonequilibrium density matrix $\rho_{ij}(t)$, which
encodes nonlocal electronic correlations across the system.

An interesting insight is obtained by expressing the electronic state in the
instantaneous eigenbasis of the single-particle Hamiltonian
$h(\{Q_i(t)\})$. Denoting its eigenvectors by $\phi^{(n)}_{i}(t)$, we define the
instantaneous population of the $n$-th eigenmode or band as
\begin{align}
f_n(t) = \mathrm{Tr}\!\left[\tilde{\rho}_n(t)\,\rho(t)\right],
\end{align}
where $\tilde{\rho}$ is the projection operator onto the $n$-th band. In terms of lattice site basis, it is explicitly given by the eivenvectors as
\begin{eqnarray}
    \tilde{\rho}_n(t) = \sum_{ij} \phi^{(n)\,*}_{i}(t)\, \phi^{(n)}_{j}(t).
\end{eqnarray}
In equilibrium, $f_n$ reduces to the Fermi--Dirac distribution. Out of equilibrium,
$f_n(t)$ can deviate strongly from this form, reflecting the presence of nonthermal
electronic excitations. Nevertheless, one can construct an adiabatic reference
density,
\begin{equation}
n_i^{\mathrm{ad}}(t)=\sum_n \bigl|\phi^{(n)}_{i}(t)\bigr|^2\, f_n(t),
\end{equation}
which depends on the instantaneous lattice configuration through the eigenstates
$\phi_{in}(t)$ and on the band populations $f_n(t)$.

As shown in Fig.~\ref{fig:decomp}, the adiabatic density $n_i^{\mathrm{ad}}(t)$
closely follows the fully nonadiabatic density $n_i^{\mathrm{nad}}(t)$ at both
early and late times. The difference
\begin{equation}
\chi_i(t)=n_i^{\mathrm{nad}}(t)-n_i^{\mathrm{ad}}(t)
\end{equation}
remains small, with fluctuations typically below $\sim 10\%$ of the average
density, as illustrated by the inset distributions. This indicates that the force acting on the lattice can be naturally decomposed into a smooth, slow adiabatic contribution determined by the instantaneous lattice configuration, supplemented by comparatively small fast fluctuating corrections.

Motivated by this observation, we model the nonadiabatic correction in a
Langevin-like form,
\begin{equation}
g\,\chi_i = -\gamma\,\dot{Q}_i + \eta_i,
\end{equation}
where $\gamma$ is an effective damping coefficient encoding the back-action of the
nonequilibrium electronic excitations on the lattice, and $\eta_i(t)$ is a
stochastic noise term capturing residual fluctuations beyond the adiabatic force.
The noise satisfies $\langle \eta_i(t) \rangle = 0$ and exhibits short-ranged
temporal correlations, reflecting the separation between fast electronic and
slow phononic time scales. Although the system is generically out of equilibrium
and no fluctuation--dissipation relation is assumed \emph{a priori}, this form
provides a physically transparent description of the effective lattice dynamics
emerging from the underlying electron--phonon coupling. We will discuss this in detail in Sec.~\ref{sec:recov}.

From a computational perspective, this decomposition is also significant.
Computing the adiabatic density $n_i^{\mathrm{ad}}(t)$ requires access to the
instantaneous electronic structure, while the full nonadiabatic evolution
requires explicit propagation of the density matrix. The force can therefore be
viewed as a functional mapping from the instantaneous lattice configuration to a
predominantly local slow adiabatic contribution, supplemented by relatively small
corrections capturing nonequilibrium effects.

These considerations motivate a machine-learning approach in which the
nonequilibrium force is approximated directly as a functional of the
instantaneous local lattice configuration. Such a description exploits the intrinsic
locality of the force and the observed separation between slow and
fast contributions, providing a practical surrogate for the expensive
force evaluation step while retaining the essential influence of nonequilibrium
electronic excitations on the lattice dynamics.

\begin{figure*}[!t]
\centering
\includegraphics[width=1.99\columnwidth]{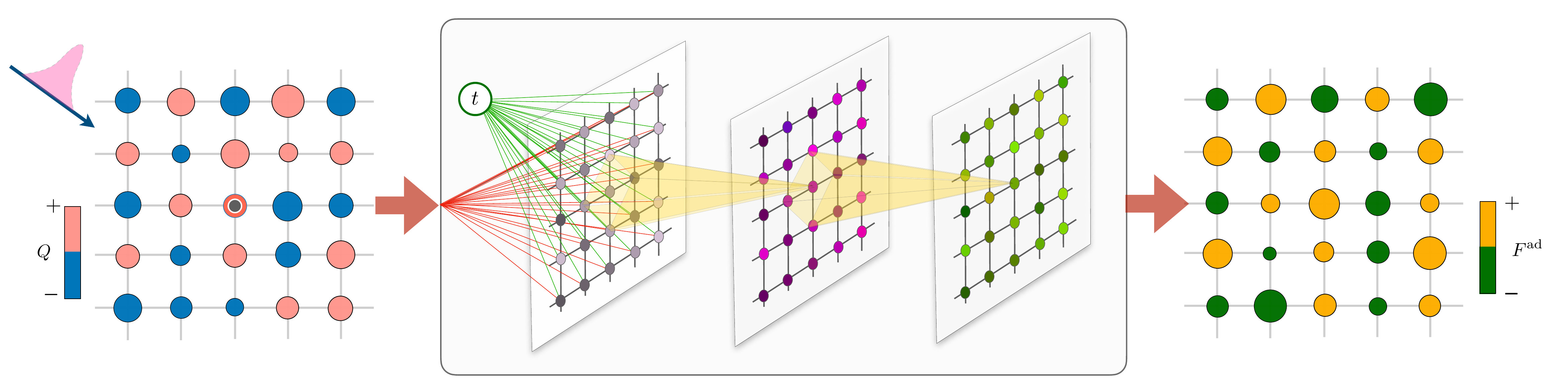}
\caption{Schematic of the graph neural network (GNN) framework for learning the adiabatic lattice force.
(Left) Snapshot of lattice distortions $\{Q_i\}$ in the two-dimensional Holstein model following excitation by a short laser pulse, illustrating a nonequilibrium distorted configuration. The set of on-site distortions $\{Q_i\}$ serves as the input to the machine-learning model. (Middle) Graph neural network architecture, where lattice sites are represented as nodes and local neighborhoods are encoded through message-passing layers. Multiple hidden layers iteratively aggregate information from nearby sites, capturing the local environment dependence implied by the locality principle. (Right) GNN output corresponding to the predicted adiabatic on-site force $F_i^{\rm ad}$ acting on each lattice site. The model thus provides a linear-scaling, symmetry-aware mapping from instantaneous lattice configurations to the adiabatic force field.
}
    \label{fig:GNN-scheme}
\end{figure*}

\section{Machine learning force-field model}
\label{sec:ML}

In this section, we introduce a machine-learning (ML) force-field framework for modeling the adiabatic component of the lattice force,
\begin{eqnarray}
	\label{eq:F_ad}
	F^{\rm ad}_i(t) = g n^{\rm ad}_i(t) = g \sum_n \bigl|\phi^{(n)}_{i}(t)\bigr|^2 f_n(t),
\end{eqnarray}
which governs the quasi-adiabatic lattice dynamics in the presence of electronically driven nonequilibrium processes. A defining advantage of ML force-field methods is their favorable scalability. At a fundamental level, linear-scaling electronic-structure approaches become viable when a system satisfies the nearsightedness principle~\cite{Kohn1996,Prodan2005}. This locality principle, rooted in wave-mechanical destructive interference, asserts that electronic properties depend predominantly on the local environment and holds broadly for both insulating and metallic systems. Exploiting this intrinsic locality is therefore essential for constructing computationally efficient models of large-scale electron-lattice dynamics.

The nearsightedness principle is naturally embedded in modern machine-learning force-field frameworks, enabling linear-scaling performance for large systems. A paradigmatic example is the neural-network force-field architecture introduced by Behler and Parrinello (BP). Originally developed in the context of quantum chemistry, the BP framework enables large-scale molecular dynamics simulations with near–first-principles accuracy by learning interatomic forces from electronic-structure calculations~\cite{behler07,bartok10,li15,shapeev16,behler16,botu17,smith17,zhang18,deringer19,mcgibbon17,suwa19,chmiela17,chmiela18,sauceda20,unke21}. In this approach, the total energy is decomposed into a sum of atomic contributions, each represented as a function of handcrafted local descriptors—such as symmetry functions—that encode the local atomic environment and explicitly enforce fundamental symmetries, including translational, rotational, and permutational invariance.

Since its introduction, the BP framework and its variants have been extended far beyond molecular systems to a broad class of condensed-matter settings, including coupled electron–lattice and electron–spin systems~\cite{zhang20,zhang21,zhang22b,zhang23,cheng23a,cheng23b,Ghosh24,Fan24,tyberg25,Jang25,Liu22,Ma19}. These developments have enabled large-scale dynamical simulations of correlated materials, opening access to emergent phenomena such as unconventional coarsening dynamics, nonequilibrium phase separation, and complex pattern formation—regimes that are typically inaccessible to direct electronic-structure approaches.

Graph neural networks (GNNs) provide an elegant and flexible alternative to BP-type force-field models, particularly well suited for lattice-based condensed-matter systems. In the GNN framework, the system is represented as a graph whose nodes correspond to lattice sites and whose edges encode local interactions or neighborhoods. Forces are learned through iterative message-passing operations that aggregate information from neighboring nodes, thereby enforcing locality through finite-range information propagation. This formulation naturally leads to linear-scaling computational cost, as the force on a given site depends only on a bounded local neighborhood.

Our objective here is to construct a GNN-based model that accurately maps a lattice distortion configuration ${ Q_i }$, provided at the input layer, to the corresponding quasi-adiabatic force components ${ F^{\rm ad}_i }$. A schematic overview of the GNN architecture is shown in Fig.~\ref{fig:GNN-scheme}. The central idea is to view the entire square lattice, subject to periodic boundary conditions, as a single graph: lattice sites are identified with graph nodes, while edges encode local connectivity defined by nearest-neighbor or extended neighbor shells. The scalar lattice distortion $Q_i$ serves as the initial node feature at site $i$.

Another distinctive feature of our GNN force-field model is the explicit inclusion of time $t$ as an input variable, with $t=0$ corresponding to the incidence of the laser pulse. This reflects the inherently nonequilibrium character of the problem: following photoexcitation, the electronic subsystem is driven far from equilibrium, and the adiabatic lattice forces are no longer uniquely determined by the instantaneous lattice configuration alone. Instead, the forces defined in Eq.~(\ref{eq:F_ad}) acquire an explicit time dependence through the evolving band occupations $f_n(t)$. This treatment contrasts with conventional machine-learning force fields constructed within the Born–Oppenheimer approximation, where electronic degrees of freedom are assumed to remain in instantaneous equilibrium and the forces are taken to be a single-valued function of atomic positions. By conditioning the force prediction on both ${Q_i}$ and $t$, our GNN effectively learns a time-dependent adiabatic force field appropriate to different stages of the nonequilibrium evolution, while preserving locality, symmetry, and linear scaling. This provides a natural extension of ML force-field approaches to driven quantum materials beyond the equilibrium Born–Oppenheimer paradigm.

The GNN is composed of multiple stacked message-passing layers, each implementing an identical local update rule at every lattice site. In each layer, information is exchanged exclusively along graph edges, such that the feature representation at a given node is updated by aggregating messages from its neighboring sites and combining them with its own current hidden state. Let $V^{(k)}_{i, \alpha}$ denote the node features associated with site $i$ at the $k$-th layer of the GNN, where $\alpha = 1, 2, \cdots, M_k$ labels the feature channels and $M_k$ is the number of channels in that layer. The message-passing update from layer $k$ to layer $k+1$ may be written in the form
\begin{eqnarray}
	\label{eq:fwd-propagation}
	U^{(k+1)}_{i, \alpha} = \sum_{\beta} W^{(k)}_{\alpha\beta} V^{(k)}_{i, \alpha} +   \sum_{j \in \mathcal{N}(i)} \sum_{\beta} \tilde{W}^{(k)}_{\alpha\beta} V^{(k)}_{j, \beta} + b^{(k)}_{\alpha}, \nonumber \\
\end{eqnarray}
where $\mathcal{N}(i)$ denotes the neighborhood of site $i$, $W^{(k)}$ and $\tilde{W}^{(k)}$ are trainable weight matrices associated with on-site and neighbor contributions, respectively, and $b^{(k)}$ is a bias vector. Nonlinearity is introduced via an element-wise activation function,
\begin{eqnarray}
	\label{eq:activation}
	V^{(k+1)}_{i, \alpha} = \mathbb{F}_{\rm av}\left(U^{(k+1)}_{i, \alpha}\right),
\end{eqnarray}
where $\mathbb{F}_{\rm av}(x)$ denotes a standard activation function such as the Rectified linear unit (ReLU).  At the input layer, the node features consist of two channels corresponding to the lattice distortion, $V^{(0)}_i = Q_i$ and the time $t$, common to all sites. At the output layer, the number of channels is again reduced to one, yielding the predicted site-resolved force $F^{\rm ad}_i$.

A key structural property of this architecture is the complete sharing of learnable parameters across all lattice sites. Specifically, the weight matrices $W^{(k)}$, $\tilde{W}^{(k)}$, and bias vectors $b^{(k)}$ are site-independent, ensuring that all nodes undergo identical computations. This parameter sharing enforces discrete translation invariance by construction: lattice sites related by translations are processed equivalently and differ only through their local environments. Moreover, the aggregation of neighbor features in the second term of Eq.~(\ref{eq:fwd-propagation}) employs the same coupling matrix $\tilde{W}^{(k)}$ for all neighboring sites $j$, thereby explicitly preserving the $D_4$ point-group symmetry of the square lattice.

As the number of message-passing layers increases, the effective receptive field associated with each node grows, allowing the network to encode progressively more extended local environments while preserving strict locality at the level of each update. Consequently, the local adiabatic force at site $i$ can be expressed as a symmetry-preserving nonlinear functional $\mathcal{F}$ of lattice distortions within a finite neighborhood,
\begin{eqnarray}
	F^{\rm ad}_i = \mathcal{F}\left( \{ Q_j \, \big| \, |\mathbf r_j - \mathbf r_i | \le r_c \} \right),
\end{eqnarray}
where the locality range $r_c$ is determined by the depth of the GNN. In this manner, the GNN provides a hierarchical and physically transparent realization of a local force field that closely mirrors the nearsightedness principle and is particularly well suited for modeling large-scale lattice dynamics, as illustrated schematically in Fig.~\ref{fig:GNN-scheme}.

\begin{table}[t]
\begin{ruledtabular}
\begin{tabular}{|c|cc|}
\textrm{Layer}&\textrm{Network}&\\
\colrule
Input (embedding) layer & \makecell[c]{GCN(2,128)\footnote{Graph convolutional layer with arguments (input size, output size).}\\act\footnote{The activation function.} =ReLU} &\\
\hline
Hidden Layer 1 & \makecell[c]{GCN(128,512)\\act =ReLU} &\\
\hline
Hidden Layer 2 & \makecell[c]{GCN(512,1024)\\act =ReLU} &\\
\hline
Hidden Layer 3 & \makecell[c]{GCN(1024,2048)\\act =ReLU} &\\
\hline
Hidden Layer 4 & \makecell[c]{GCN(2048,1024)\\act =ReLU} &\\
\hline
Hidden Layer 5 & \makecell[c]{GCN(1024,512)\\act =ReLU} &\\
\hline
Hidden Layer 6 & \makecell[c]{GCN(512,128)\\act =ReLU} &\\
\hline
Output Layer & GCN(128,1) & \\
\end{tabular}
\end{ruledtabular}
\caption{Graph neural network architecture and hyperparameters used for electronic force prediction.}
\label{tab:GCN_table}
\end{table}

The GNN force-field model is implemented using PyTorch~\cite{Paszke2019}, with the forward propagation of node features following Eqs.~(\ref{eq:fwd-propagation}) and (\ref{eq:activation}) realized through stacked graph convolutional network (GCN) layers. The network comprises a total of eight GCN layers. Each lattice site is characterized by two input features: the local lattice displacement and a temporal descriptor defined as the reciprocal of the time index associated with the corresponding snapshot. These inputs are first projected into a 128-dimensional latent space via an embedding layer, after which the features are processed through a sequence of GCN layers with widths
$128 \rightarrow 512 \rightarrow 1024 \rightarrow 2048 \rightarrow 1024 \rightarrow 512 \rightarrow 128 \rightarrow 1$,
terminating in a single-channel output layer that predicts the local adiabatic force. Separate trainable weight matrices are employed for edge-mediated message passing and self-loop (on-site) contributions, consistent with the formulation in Eq.~(\ref{eq:fwd-propagation}). Rectified linear unit (ReLU)~\cite{Nair2010} activations are used throughout the network.

\begin{figure}[t]
\centering
\includegraphics[width=\linewidth]{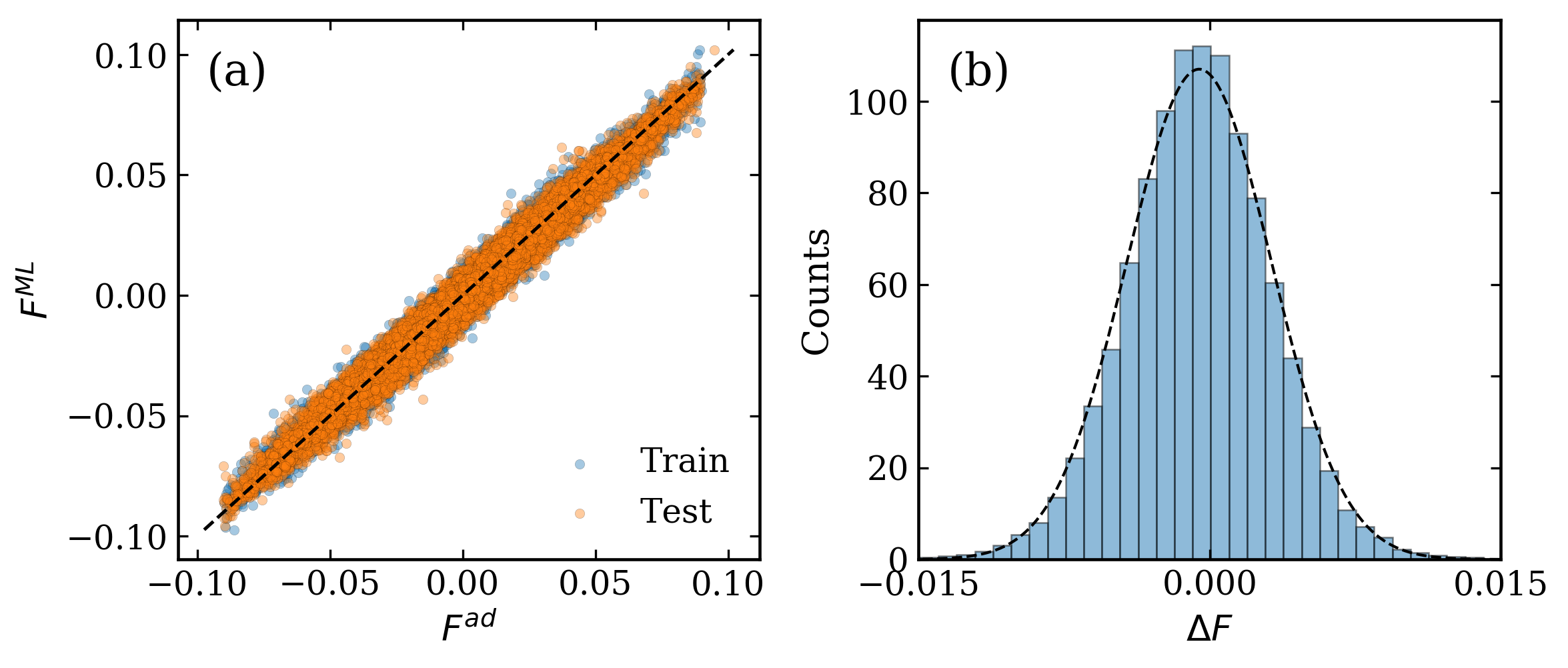}
\caption{Benchmark of the machine-learning force prediction against the exact adiabatic force. (a) Scatter plot comparing the ML-predicted force $F^{\rm ML}$ with the adiabatic force $F^{\rm ad}$ obtained from the reference calculation. The close clustering of data points along the diagonal indicates excellent agreement over the full force range. (b) Histogram of the prediction error $\Delta F = F^{\rm ML} - F^{\rm ad}$, demonstrating a narrow, approximately symmetric error distribution centered near zero.
}
    \label{fig:MLexact}
\end{figure}

The model is trained by minimizing a mean-squared-error (MSE) loss function defined over the predicted forces at selected time snapshots,
\begin{eqnarray}
	\mathcal{L} = \sum_{t} \sum_i \left| F^{\rm ad}_i(t) - F^{\rm ML}_i(t) \right|^2,
\end{eqnarray}
which penalizes deviations between the GNN-predicted forces and the exact adiabatic forces at each lattice site. This choice of loss function directly reflects the physical objective of accurately reproducing the local force field governing the lattice dynamics, rather than intermediate quantities such as energies or densities. By summing over both spatial sites and time snapshots, the loss enforces consistency across different stages of the nonequilibrium evolution and encourages the model to learn a force field that generalizes throughout the recovery dynamics. The model parameters are optimized using the Adam optimizer~\cite{Kingma2017} with a learning rate of 0.001. The training and validation datasets are constructed from lattice configurations drawn from 30 independent nonequilibrium dynamical runs, spanning time steps from 2500 to 20,000 and thereby providing broad coverage of the recovery dynamics. Of these trajectories, 28 runs are used for training, while the remaining two are reserved for validation. A detailed summary of the GCN architecture and associated hyperparameters is provided in Table~\ref{tab:GCN_table}.

Fig.~\ref{fig:MLexact} benchmarks the performance of the GNN force-field model against the exact adiabatic forces. As shown in panel (a), the ML-predicted forces $F^{\rm ML}$ exhibit an excellent one-to-one correspondence with the reference adiabatic forces $F^{\rm ad}$ over the entire force range, with both training and test data tightly clustered around the diagonal. The near overlap of the training and test distributions indicates good generalization and the absence of overfitting, despite the highly nonequilibrium nature of the underlying dynamics. Panel (b) further quantifies the accuracy through the distribution of force errors $\Delta F = F^{\rm ML} - F^{\rm ad}$. The error histogram is narrowly peaked around zero and well described by an approximately Gaussian profile, demonstrating that the model achieves high precision with negligible systematic bias. Together, these results confirm that the GNN reliably captures the local adiabatic force landscape across different configurations and times, providing a robust and transferable force-field representation for nonequilibrium lattice dynamics.

\section{Dynamics of Charge-Density-Wave Recovery}
\label{sec:recov}
Having trained a model that predicts the electronic force on site $i$ from the instantaneous background configuration $\{Q_i\}$, we now construct the equation of motion for the lattice degrees of freedom. The full dynamical equation is given by
\begin{equation}
m\ddot{Q}_i = -m\Omega^2 Q_i + F_i^{\rm ML}(\{Q_i\}) + \chi_i(t),
\end{equation}
where $F_i^{\rm ML}$ denotes the adiabatic electronic force obtained from the trained model and $\chi_i(t)$ can be interpreted as the back-action of the electronic subsystem on the lattice degrees of freedom, treated as an effective bath.

Although the trained model accurately captures the electronic force during both the pump-induced melting and the subsequent recovery of the ordered state, the bath response during the strongly nonequilibrium melting regime is more difficult to model reliably due to the rapid redistribution of electronic populations and the presence of strong transient currents. For this reason, in the present study we focus on the recovery dynamics, where the electronic subsystem relaxes more gradually and an effective bath description is better justified. We initialize the recovery stage at time $t_1 = 80\tau_0$ and use the lattice configuration $\{Q_i(t_1)\}$ obtained from the exact nonadiabatic simulation as the starting point for all benchmark comparisons.

Following the approach discussed in Ref.~\cite{HolstPhoto2}, the origin of this bath term can be understood by noting that although the total energy of the coupled electron--lattice system is conserved after photoexcitation, our description explicitly focuses on the slow phonon variables while integrating out the fast electronic degrees of freedom. This procedure generates an effective bath response acting on the lattice dynamics. In the equilibrium insulating phase, the absence of low-energy electronic states near the Fermi level suppresses dissipation channels, resulting in weakly damped phonon motion. In contrast, photoexcitation creates a finite population of carriers across the gap, injects energy into the electronic sector, and generates transient currents, which open additional scattering channels and enhance lattice damping. As the electronic subsystem gradually relaxes and redistributes energy among electronic and lattice degrees of freedom, this excess dissipation correspondingly decays. In general, the resulting bath response is non-Markovian and contains memory effects. However, in the adiabatic regime, where the phonon bandwidth is much smaller than the electronic bandwidth, the electronic subsystem relaxes on much shorter timescales. In this wide-band limit, the bath kernel becomes local in time, leading, to a good approximation, to a Markovian Langevin description with Gaussian white noise~\cite{Sclass_holst5,Lang1}.

\begin{figure}[t]
\centering
\includegraphics[width=0.95\linewidth]{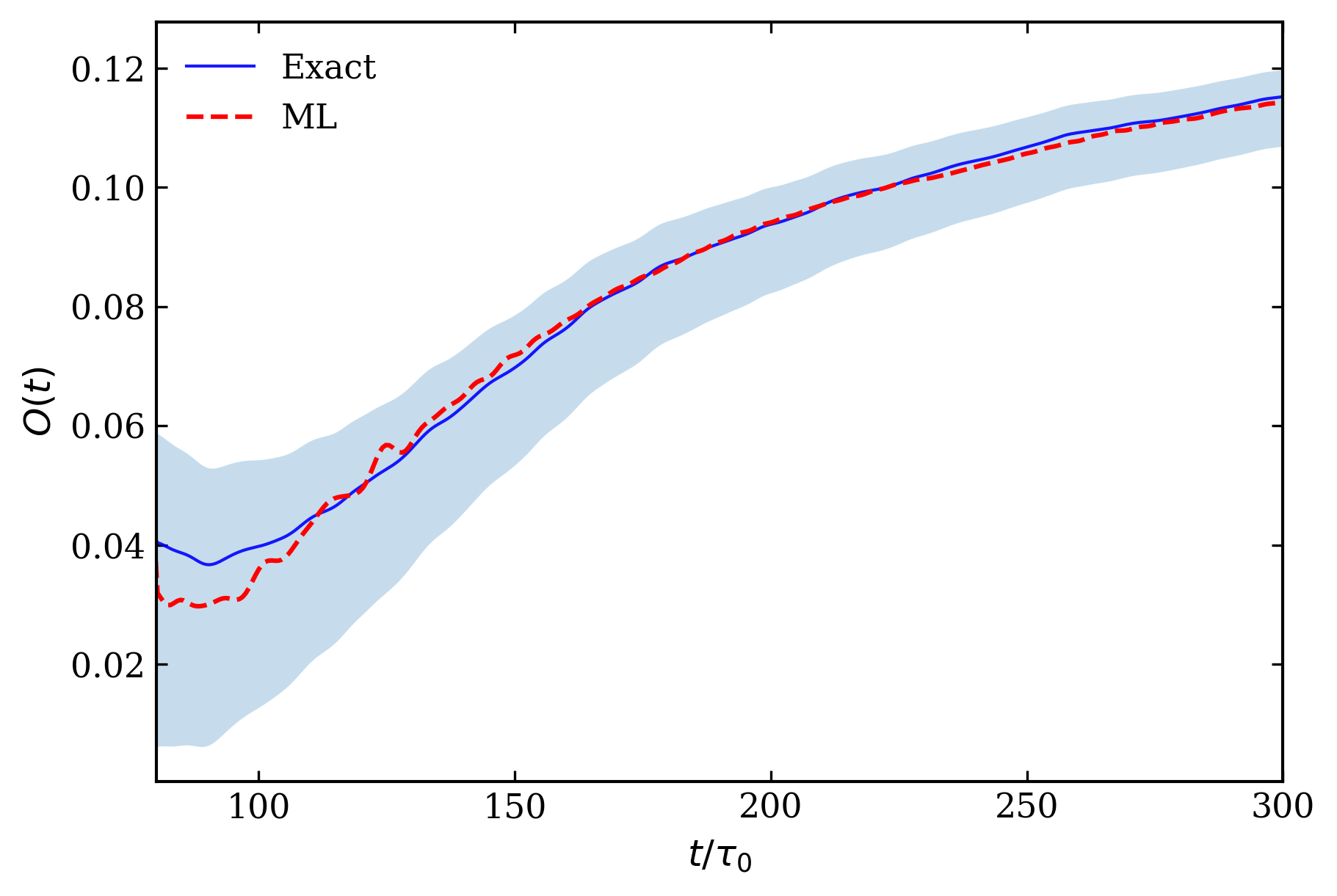}
\caption{%
Recovery dynamics of the order parameter following photoexcitation.
The exact nonadiabatic evolution (black solid line) is ensemble averaged over 30 independent initial conditions, with the shaded band indicating the corresponding standard deviation.
The ML-based dynamics (red solid line) are averaged over the same number of realizations.
The close agreement between the two curves, together with the overlap within the fluctuation window, demonstrates the accuracy of the learned force model combined with the effective bath description.
}
    \label{fig:traj}
\end{figure}

We therefore parametrize the bath contribution as
\begin{equation}
\chi_i(t) = -\gamma(t)\dot{Q}_i + \eta_i(t),
\end{equation}
where $\gamma(t)$ is the dissipation rate and $\eta_i(t)$ is a stochastic noise term satisfying
\begin{equation}
\langle \eta_i(t)\eta_j(t') \rangle = 2\gamma(t)T\,\delta_{ij}\delta(t-t').
\end{equation}
The effective bath temperature $T$ is defined operationally from the long-time velocity distribution using the equipartition theorem
\begin{equation}
\langle m\dot{Q}_i^2 \rangle = T,
\end{equation}
and is subsequently treated as a constant throughout the evolution, capturing the average energy transferred from the electronic subsystem to the lattice sector.

In contrast, the dissipation strength generally retains explicit time dependence, reflecting the nonequilibrium electronic population created by the pump pulse. Immediately after photoexcitation, the presence of excited carriers enhances electronic scattering channels and increases lattice damping. As the electronic subsystem relaxes and the photoexcited carrier population decays, the associated excess dissipation is expected to diminish. To capture this behavior at a minimal phenomenological level, we model the time-dependent damping as
\begin{equation}
\gamma(t) = \gamma_0 + \Delta\gamma \, e^{-(t-t_1)/\tau_\gamma},
\end{equation}
where $\gamma_0$ denotes the steady-state background dissipation, $\Delta\gamma$ is the pump-induced enhancement of the damping, and $\tau_\gamma$ is the characteristic electronic relaxation timescale.

\begin{figure}[t]
\centering
\includegraphics[width=0.9\linewidth, height=11cm]{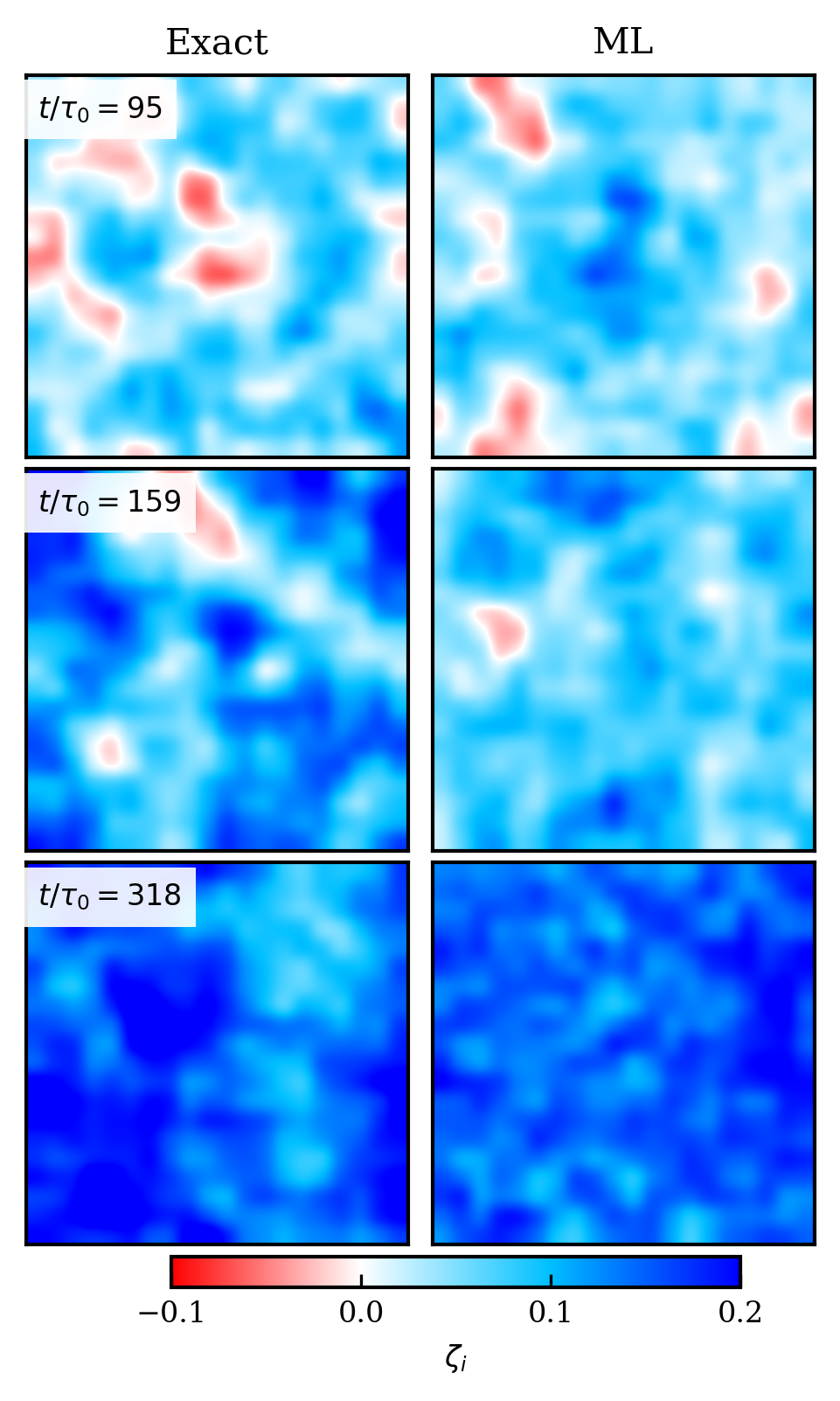}
\caption{%
Real-space map of local lattice CDW order $\zeta_i$ at three representative times corresponding to distinct stages of the recovery dynamics for a single stochastic realization.
The left column shows results from the exact nonadiabatic evolution, while the right column shows the corresponding ML-based dynamics.
}
    \label{fig:config}
\end{figure}

Using the parameter values $\gamma_0 = 0.003$, $\Delta\gamma = 0.017$, and $\tau_\gamma = 25\tau_0$, we obtain excellent agreement with the recovery dynamics of the order parameter, as shown in Fig.~\ref{fig:traj}. In this comparison, the exact nonadiabatic evolution is ensemble averaged over 30 independent initial conditions, with a shaded envelope indicating the corresponding standard deviation. The machine-learning-based dynamics are averaged over the same number of realizations and are found to lie well within the fluctuation envelope of the exact results.
We further compare real-space snapshots of the local CDW order parameter, 
\begin{equation}
\zeta_i = (Q_i - Q_0) e^{i \textbf{K}\cdot \textbf{R}_i} 
\end{equation} 
where $Q_0=g/2m\Omega^2$, obtained from the exact nonadiabatic and ML-based schemes at three representative times corresponding to distinct stages of the recovery dynamics, as shown in Fig.~\ref{fig:config}, for a single stochastic realization. Despite the inherent stochasticity of the dynamics, the ML results closely reproduce the spatial patterns and domain structures of the exact evolution, demonstrating strong qualitative and quantitative agreement.

\section{Conclusion and outlook}
\label{sec:conclusion}

In equilibrium and near-adiabatic regimes, lattice dynamics in electron--phonon systems such as the Holstein model is commonly described by forces evaluated from the instantaneous lattice configuration, reflecting the adiabatic response of the electronic subsystem. In this work, we have shown that a closely related physical picture remains applicable even far from equilibrium. Focusing on the laser-driven Holstein model, we demonstrated that after strong photoexcitation the nonequilibrium electron--lattice dynamics naturally separates into slow lattice motion and fast electronic processes, once the effects of the nonequilibrium carrier population are properly accounted for. The resulting lattice force can be expressed as the sum of a smooth, slowly varying quasi-adiabatic contribution determined by the instantaneous lattice configuration and a smaller, rapidly fluctuating term arising from fast electronic dynamics, which effectively acts as a bath for the slow lattice order parameters.

Building on this separation, we showed that the dominant quasi-adiabatic force remains local and well behaved throughout the recovery regime and can therefore be learned efficiently using machine-learning techniques. By training a graph neural network to represent this force field, we achieved linear-scaling evaluation of lattice forces while maintaining quantitative agreement with exact adiabatic reference calculations. When combined with a minimal Langevin description of the bath response, the resulting ML-based dynamics accurately reproduces both the recovery of the global CDW order parameter and the evolution of real-space domain structures obtained from fully nonadiabatic simulations. This establishes a practical route for substantially reducing the computational cost of nonequilibrium simulations by replacing explicit electronic time evolution with a data-driven surrogate description of the slow sector.

The present study provides a proof-of-principle implementation of this framework. Here, we focused on benchmarking the quasi-adiabatic force component and adopted a phenomenological time-dependent dissipation model to describe the bath response during recovery. More refined treatments are possible and represent natural directions for future work, including incorporating non-Markovian memory effects, learning the bath response directly from microscopic data, extending the ML framework to jointly model force and noise correlations, and explicitly incorporating time-dependent electronic observables such as carrier populations and currents as auxiliary inputs. Such extensions will be important for capturing the strongly nonequilibrium melting regime and ultrafast transient dynamics immediately following photoexcitation.

More broadly, our results indicate that the slow--fast separation identified here is not specific to the Holstein model, but reflects a generic feature of nonequilibrium systems in which slow collective degrees of freedom are coupled to fast fermionic excitations. Combined with machine-learning force-field techniques, this perspective provides a general strategy for simulating driven quantum materials, enabling scalable access to long-time dynamics in regimes where direct microscopic approaches remain computationally prohibitive. This framework can be extended to a wide range of problems involving electron--phonon, spin--fermion, and multi-orbital coupled systems, opening new opportunities for studying emergent nonequilibrium phenomena in correlated materials.

\vspace{-0.2cm}
\begin{acknowledgments}

This work was supported by the Owens Family Foundation. Y.F. was partially supported by the US Department of Energy Basic Energy Sciences under Contract No. DE-SC0020330. The authors acknowledge Research Computing at the University of Virginia for providing computational resources and technical support.
\end{acknowledgments}

\bibliography{ref,copp}

@article{OS1,
  author  = {Rohwer, T. and Hellmann, S. and Wiesenmayer, M.},
  title   = {Collapse of long-range charge order tracked by time-resolved photoemission at high momenta},
  journal = {Nature},
  volume  = {471},
  pages   = {490--493},
  year    = {2011},
  doi     = {10.1038/nature09829}
}

@article{OS2,
  author  = {Okamoto, H. and Miyagoe, T. and Kobayashi, K. and Uemura, H. and Nishioka, H. and Matsuzaki, H. and Sawa, A. and Tokura, Y.},
  title   = {Photoinduced transition from Mott insulator to metal in the undoped cuprates Nd$_2$CuO$_4$ and La$_2$CuO$_4$},
  journal = {Phys. Rev. B},
  volume  = {83},
  pages   = {125102},
  year    = {2011},
  doi     = {10.1103/PhysRevB.83.125102}
}

@article{OS3,
  author  = {Beaud, P. and Caviezel, A. and Mariager, S.},
  title   = {A time-dependent order parameter for ultrafast photoinduced phase transitions},
  journal = {Nature Materials},
  volume  = {13},
  pages   = {923--927},
  year    = {2014},
  doi     = {10.1038/nmat4046}
}

@article{OS4,
  author  = {Tomimoto, S. and Miyasaka, S. and Ogasawara, T. and Okamoto, H. and Tokura, Y.},
  title   = {Ultrafast photoinduced melting of orbital order in LaVO$_3$},
  journal = {Phys. Rev. B},
  volume  = {68},
  pages   = {035106},
  year    = {2003},
  doi     = {10.1103/PhysRevB.68.035106}
}

@article{OS5,
  author  = {Tomeljak, A.},
  title   = {Dynamics of photoinduced charge-density-wave to metal phase transition in K$_{0.3}$MoO$_3$},
  journal = {Phys. Rev. Lett.},
  volume  = {102},
  pages   = {066404},
  year    = {2009},
  doi     = {10.1103/PhysRevLett.102.066404}
}

@article{PICO1,
  author  = {Lee, W. and Chuang, Y. and Moore, R.},
  title   = {Phase fluctuations and the absence of topological defects in a photo-excited charge-ordered nickelate},
  journal = {Nature Communications},
  volume  = {3},
  pages   = {838},
  year    = {2012},
  doi     = {10.1038/ncomms1837}
}

@article{PICO2,
  author  = {Kogar, A.},
  title   = {Light-induced charge density wave in LaTe$_3$},
  journal = {Nature Physics},
  volume  = {16},
  pages   = {159--163},
  year    = {2020},
  doi     = {10.1038/s41567-019-0705-3}
}

@article{PICO3,
  author  = {Ravnik, J. and Diego, M. and Gerasimenko, Y.},
  title   = {A time-domain phase diagram of metastable states in a charge ordered quantum material},
  journal = {Nature Communications},
  volume  = {12},
  pages   = {2323},
  year    = {2021},
  doi     = {10.1038/s41467-021-22646-7}
}

@article{PICO4,
  author  = {Ravnik, J. and Vaskivskyi, I. and Mertelj, T. and Mihailovic, D.},
  title   = {Real-time observation of the coherent transition to a metastable emergent state in 1T-TaS$_2$},
  journal = {Phys. Rev. B},
  volume  = {97},
  pages   = {075304},
  year    = {2018},
  doi     = {10.1103/PhysRevB.97.075304}
}

@article{PICO5,
  author  = {Vaskivskyi, I.},
  title   = {Fast electronic resistance switching involving hidden charge density wave states},
  journal = {Nature Communications},
  volume  = {7},
  pages   = {11442},
  year    = {2016}
}

@article{PICO6,
  author  = {Yoshikawa, N. and Suganuma, H. and Matsuoka, H. and Tanaka, Y. and Hemme, P. and Cazayous, M. and Gallais, Y. and Nakano, M. and Iwasa, Y. and Shimano, R.},
  title   = {Ultrafast switching to an insulating-like metastable state by amplitudon excitation of a charge density wave},
  journal = {Nature Physics},
  volume  = {17},
  pages   = {909--914},
  year    = {2021},
  doi     = {10.1038/s41567-021-01267-3}
}

@article{Holst2,
  author  = {Holstein, T.},
  title   = {Studies of polaron motion: Part II. The small polaron},
  journal = {Annals of Physics},
  volume  = {8},
  pages   = {343--389},
  year    = {1959},
  doi     = {10.1016/0003-4916(59)90003-X}
}

@article{Holst3,
  author  = {Fehske, H. and Loos, J. and Wellein, G.},
  title   = {Quantum lattice dynamical effects on the Peierls transition in the Holstein model},
  journal = {Phys. Rev. B},
  volume  = {61},
  pages   = {8016--8022},
  year    = {2000},
  doi     = {10.1103/PhysRevB.61.8016}
}

@article{Sclass_holst1,
  author  = {Kumar, Sanjeev and Majumdar, Pinaki},
  title   = {Charge ordering and polaron formation in the adiabatic Holstein model},
  journal = {Phys. Rev. Lett.},
  volume  = {91},
  pages   = {246602},
  year    = {2003},
  doi     = {10.1103/PhysRevLett.91.246602}
}

@article{Sclass_holst2,
  author  = {Kumar, Sanjeev and Majumdar, Pinaki},
  title   = {Phase diagram of the adiabatic Holstein model: Charge order, polarons, and metal-insulator transitions},
  journal = {Phys. Rev. B},
  volume  = {68},
  pages   = {115109},
  year    = {2003},
  doi     = {10.1103/PhysRevB.68.115109}
}

@article{Sclass_holst3,
  author  = {Poornachandra Sekhar, B. and Kumar, Sanjeev and Majumdar, Pinaki},
  title   = {Many-electron ground state of the adiabatic Holstein model in two and three dimensions},
  journal = {Phys. Rev. B},
  volume  = {72},
  pages   = {075102},
  year    = {2005},
  doi     = {10.1103/PhysRevB.72.075102}
}

@article{Sclass_holst4,
  author  = {Kumar, Sanjeev and Majumdar, Pinaki},
  title   = {Static and transport properties of the adiabatic Holstein model},
  journal = {Eur. Phys. J. B},
  volume  = {50},
  pages   = {571--579},
  year    = {2006},
  doi     = {10.1140/epjb/e2006-00154-2}
}

@article{Sclass_holst5,
  title = {Langevin approach to lattice dynamics in a charge-ordered polaronic system},
  author = {Bhattacharyya, Sauri and Bakshi, Sankha Subhra and Kadge, Samrat and Majumdar, Pinaki},
  journal = {Phys. Rev. B},
  volume = {99},
  issue = {16},
  pages = {165150},
  numpages = {16},
  year = {2019},
  month = {Apr},
  publisher = {American Physical Society},
  doi = {10.1103/PhysRevB.99.165150},
  url = {https://link.aps.org/doi/10.1103/PhysRevB.99.165150}
}

@misc{ML2,
  author       = {Yang, Yang and Cheng, Chen and Fan, Yunhao and Chern, Gia-Wei},
  title        = {Enhanced coarsening of charge density waves induced by electron correlation: Machine-learning enabled large-scale dynamical simulations},
  howpublished = {arXiv:2412.21072 [cond-mat.stat-mech]},
  year         = {2024}
}

@article{ML3,
  author  = {Cheng, Chen and Zhang, Sheng and Fan, Yunhao and Chern, Gia-Wei},
  title   = {Kinetics of Peierls dimerization transition: Machine learning force-field approach},
  journal = {arXiv:2510.20659 [cond-mat.str-el]},
  year    = {2025}
}

@article{ML4,
  author  = {Ghosh, Supriyo and Zhang, Sheng and Cheng, Chen and Chern, Gia-Wei},
  title   = {Kinetics of orbital ordering in cooperative Jahn-Teller models: Machine-learning enabled large-scale simulations},
  journal = {arXiv:2405.14776 [cond-mat.str-el]},
  year    = {2024}
}

@article{HolstPhoto1,
  title   = {Nonequilibrium sum rules for the Holstein model},
  author  = {Freericks, J. K. and Najafi, K. and Kemper, A. F. and Devereaux, T. P.},
  journal = {Phys. Rev. B},
  volume  = {89},
  pages   = {075139},
  year    = {2014},
  doi     = {10.1103/PhysRevB.89.075139}
}

@article{HolstPhoto2,
  title     = {Nonequilibrium dynamics of suppression, revival, and loss of charge order in a laser-pumped electron-phonon system},
  author    = {Bakshi, Sankha Subhra and Bose, Debraj and Dutta, Arijit and Majumdar, Pinaki},
  journal   = {Phys. Rev. B},
  volume    = {110},
  issue     = {7},
  pages     = {075102},
  year      = {2024},
  month     = {Aug},
  publisher = {American Physical Society},
  doi       = {10.1103/PhysRevB.110.075102}
}

@article{HolstPhoto3,
  title   = {Photoinduced pattern formation and melting of charge-density-wave order in electron-phonon systems},
  author  = {Yang, Lingyu and Chern, Gia-Wei},
  journal = {Phys. Rev. B},
  volume  = {109},
  pages   = {L121101},
  year    = {2024},
  doi     = {10.1103/PhysRevB.109.L121101}
}

@article{HolstPhoto4,
  author  = {Yang, Lingyu and Jang, Ho and Bakshi, Sankha Subhra and Yang, Yang and Chern, Gia-Wei},
  title   = {Pseudospin formulation of quench dynamics in the semiclassical Holstein model},
  journal = {arXiv:2601.01694},
  year    = {2026},
}

@article{HolstPhoto5,
  author  = {Picano, A.},
  title   = {Stochastic semiclassical theory for nonequilibrium electron-phonon systems},
  journal = {Phys. Rev. B},
  volume  = {108},
  pages   = {035115},
  year    = {2023},
  doi     = {10.1103/PhysRevB.108.035115}
}

@article{HolstPhoto6,
  title = {Pattern formation in charge density wave states after a quantum quench},
  author = {Yang, Lingyu and Yang, Yang and Chern, Gia-Wei},
  journal = {Phys. Rev. B},
  volume = {109},
  issue = {19},
  pages = {195133},
  numpages = {13},
  year = {2024},
  month = {May},
  publisher = {American Physical Society},
  doi = {10.1103/PhysRevB.109.195133},
  url = {https://link.aps.org/doi/10.1103/PhysRevB.109.195133}
}

@article{Lang1,
  title = {Intermittent polaron dynamics: Born-Oppenheimer approximation out of equilibrium},
  author = {Mozyrsky, D. and Hastings, M. B. and Martin, I.},
  journal = {Phys. Rev. B},
  volume = {73},
  issue = {3},
  pages = {035104},
  numpages = {6},
  year = {2006},
  month = {Jan},
  publisher = {American Physical Society},
  doi = {10.1103/PhysRevB.73.035104},
  url = {https://link.aps.org/doi/10.1103/PhysRevB.73.035104}
}

@article{Nowadnick12,
  title = {Competition Between Antiferromagnetic and Charge-Density-Wave Order in the Half-Filled Hubbard-Holstein Model},
  author = {Nowadnick, E. A. and Johnston, S. and Moritz, B. and Scalettar, R. T. and Devereaux, T. P.},
  journal = {Phys. Rev. Lett.},
  volume = {109},
  issue = {24},
  pages = {246404},
  numpages = {5},
  year = {2012},
  month = {Dec},
  publisher = {American Physical Society},
  doi = {10.1103/PhysRevLett.109.246404},
  url = {https://link.aps.org/doi/10.1103/PhysRevLett.109.246404}
}

@article{Costa2020,
    title = {Phase diagram of the two-dimensional Hubbard-Holstein model},
    author = {Costa, N.C. and Seki, K. and Yunoki, S. and Sorella, S.},
    journal = {Commun Phys},
    volume = {3},
    pages = {80},
    year = {2020},
    publisher = {Nature Publishing Group UK London},
    doi = {10.1038/s42005-020-0342-2},
    url = {https://doi.org/10.1038/s42005-020-0342-2}
}

@article{Johnston2013,
    title = {Determinant quantum Monte Carlo study of the two-dimensional single-band Hubbard-Holstein model},
    author = {Johnston, S. and Nowadnick, E. A. and Kung, Y. F. and Moritz, B. and Scalettar, R. T. and Devereaux, T. P.},
    journal = {Phys. Rev. B},
    volume = {87},
    issue = {23},
    pages = {235133},
    numpages = {13},
    year = {2013},
    month = {Jun},
    publisher = {American Physical Society},
    doi = {10.1103/PhysRevB.87.235133},
    url = {https://link.aps.org/doi/10.1103/PhysRevB.87.235133}
}

@article{golez12,
  title = {Relaxation Dynamics of the Holstein Polaron},
  author = {Gole\ifmmode \check{z}\else \v{z}\fi{}, Denis and Bon\ifmmode \check{c}\else \v{c}\fi{}a, Janez and Vidmar, Lev and Trugman, Stuart A.},
  journal = {Phys. Rev. Lett.},
  volume = {109},
  issue = {23},
  pages = {236402},
  numpages = {5},
  year = {2012},
  month = {Dec},
  publisher = {American Physical Society},
  doi = {10.1103/PhysRevLett.109.236402},
  url = {https://link.aps.org/doi/10.1103/PhysRevLett.109.236402}
}

@article{Kohn1996,
    title = {Density Functional and Density Matrix Method Scaling Linearly with the Number of Atoms},
    author = {Kohn, W.},
    journal = {Phys. Rev. Lett.},
    volume = {76},
    issue = {17},
    pages = {3168--3171},
    numpages = {0},
    year = {1996},
    month = {Apr},
    publisher = {American Physical Society},
    doi = {10.1103/PhysRevLett.76.3168},
    url = {https://link.aps.org/doi/10.1103/PhysRevLett.76.3168}
}

@article{Prodan2005,
    title = {Nearsightedness of electronic matter},
    author = {E. Prodan  and W. Kohn},
    journal = {Proc. Natl. Acad. Sci. USA},
    volume = {102},
    number = {33},
    pages = {11635-11638},
    year = {2005},
    doi = {10.1073/pnas.0505436102},
    url = {https://www.pnas.org/doi/abs/10.1073/pnas.0505436102}
}

@book{Marx2009, 
    title={Ab Initio Molecular Dynamics: Basic Theory and Advanced Methods}, 
    author={Marx, Dominik and Hutter, Jürg}, 
    publisher={Cambridge University Press}, 
    place={Cambridge}, 
    year={2009},
    doi={10.1017/CBO9780511609633},
    url={https://doi.org/10.1017/CBO9780511609633}
}

@article{behler07,
  title = {Generalized Neural-Network Representation of High-Dimensional Potential-Energy Surfaces},
  author = {Behler, J\"org and Parrinello, Michele},
  journal = {Phys. Rev. Lett.},
  volume = {98},
  issue = {14},
  pages = {146401},
  numpages = {4},
  year = {2007},
  month = {Apr},
  publisher = {American Physical Society},
  doi = {10.1103/PhysRevLett.98.146401},
  url = {https://link.aps.org/doi/10.1103/PhysRevLett.98.146401}
}

@article{bartok10,
  title = {Gaussian Approximation Potentials: The Accuracy of Quantum Mechanics, without the Electrons},
  author = {Bart\'ok, Albert P. and Payne, Mike C. and Kondor, Risi and Cs\'anyi, G\'abor},
  journal = {Phys. Rev. Lett.},
  volume = {104},
  issue = {13},
  pages = {136403},
  numpages = {4},
  year = {2010},
  month = {Apr},
  publisher = {American Physical Society},
  doi = {10.1103/PhysRevLett.104.136403},
  url = {https://link.aps.org/doi/10.1103/PhysRevLett.104.136403}
}

@article{li15,
  title = {Molecular Dynamics with On-the-Fly Machine Learning of Quantum-Mechanical Forces},
  author = {Li, Zhenwei and Kermode, James R. and De Vita, Alessandro},
  journal = {Phys. Rev. Lett.},
  volume = {114},
  issue = {9},
  pages = {096405},
  numpages = {5},
  year = {2015},
  month = {Mar},
  publisher = {American Physical Society},
  doi = {10.1103/PhysRevLett.114.096405},
  url = {https://link.aps.org/doi/10.1103/PhysRevLett.114.096405}
}

@Article{botu17,
author={Botu, V.
and Batra, R.
and Chapman, J.
and Ramprasad, R.},
title={Machine Learning Force Fields: Construction, Validation, and Outlook},
journal={The Journal of Physical Chemistry C},
year={2017},
month={Jan},
day={12},
publisher={American Chemical Society},
volume={121},
number={1},
pages={511-522},
issn={1932-7447},
doi={10.1021/acs.jpcc.6b10908},
url={https://doi.org/10.1021/acs.jpcc.6b10908}
}

@Article{smith17,
author ="Smith, J. S. and Isayev, O. and Roitberg, A. E.",
title  ="ANI-1: an extensible neural network potential with DFT accuracy at force field computational cost",
journal  ="Chem. Sci.",
year  ="2017",
volume  ="8",
issue  ="4",
pages  ="3192-3203",
publisher  ="The Royal Society of Chemistry",
doi  ="10.1039/C6SC05720A",
url  ="http://dx.doi.org/10.1039/C6SC05720A"
}

@article{zhang18,
  title = {Deep Potential Molecular Dynamics: A Scalable Model with the Accuracy of Quantum Mechanics},
  author = {Zhang, Linfeng and Han, Jiequn and Wang, Han and Car, Roberto and E, Weinan},
  journal = {Phys. Rev. Lett.},
  volume = {120},
  issue = {14},
  pages = {143001},
  numpages = {6},
  year = {2018},
  month = {Apr},
  publisher = {American Physical Society},
  doi = {10.1103/PhysRevLett.120.143001},
  url = {https://link.aps.org/doi/10.1103/PhysRevLett.120.143001}
}

@article{behler16,
    author = {Behler, J},
    title = "{Perspective: Machine learning potentials for atomistic simulations}",
    journal = {The Journal of Chemical Physics},
    volume = {145},
    number = {17},
    pages = {170901},
    year = {2016},
    month = {11},
    issn = {0021-9606},
    doi = {10.1063/1.4966192},
    url = {https://doi.org/10.1063/1.4966192},
}

@article{shapeev16,
author = {Shapeev, Alexander V.},
title = {Moment Tensor Potentials: A Class of Systematically Improvable Interatomic Potentials},
journal = {Multiscale Modeling \& Simulation},
volume = {14},
number = {3},
pages = {1153-1173},
year = {2016},
doi = {10.1137/15M1054183},
URL = {https://doi.org/10.1137/15M1054183},
}

@article{deringer19,
author = {Deringer, V. L. and Caro, M. A. and Csanyi, G},
title = {Machine Learning Interatomic Potentials as Emerging Tools for Materials Science},
journal = {Advanced Materials},
volume = {31},
number = {46},
pages = {1902765},
doi = {https://doi.org/10.1002/adma.201902765},
url = {https://onlinelibrary.wiley.com/doi/abs/10.1002/adma.201902765},
year = {2019}
}

@article{mcgibbon17,
	author = {McGibbon, R. T. and Taube, A. G. and Donchev, A. G. and Siva, K. and Hernandez, F. and Hargus, C. and Law, K.-H. and Klepeis, J. L. and Shaw, D. E.},
    title = {Improving the accuracy of M\"oller-Plesset perturbation theory with neural networks},
    journal = {The Journal of Chemical Physics},
    volume = {147},
    number = {16},
    pages = {161725},
    year = {2017},
    month = {09},
    issn = {0021-9606},
    doi = {10.1063/1.4986081},
    url = {https://doi.org/10.1063/1.4986081},
}

@article{suwa19,
  title = {Machine learning for molecular dynamics with strongly correlated electrons},
  author = {Suwa, Hidemaro and Smith, Justin S. and Lubbers, Nicholas and Batista, Cristian D. and Chern, Gia-Wei and Barros, Kipton},
  journal = {Phys. Rev. B},
  volume = {99},
  issue = {16},
  pages = {161107},
  numpages = {5},
  year = {2019},
  month = {Apr},
  publisher = {American Physical Society},
  doi = {10.1103/PhysRevB.99.161107},
  url = {https://link.aps.org/doi/10.1103/PhysRevB.99.161107}
}

@article{chmiela17,
author = {Stefan Chmiela  and Alexandre Tkatchenko  and Huziel E. Sauceda  and Igor Poltavsky  and Kristof T. Schütt  and Klaus-Robert Müller },
title = {Machine learning of accurate energy-conserving molecular force fields},
journal = {Science Advances},
volume = {3},
number = {5},
pages = {e1603015},
year = {2017},
doi = {10.1126/sciadv.1603015},
URL = {https://www.science.org/doi/abs/10.1126/sciadv.1603015}}

@Article{chmiela18,
author={Chmiela, Stefan
and Sauceda, Huziel E.
and M{\"u}ller, Klaus-Robert
and Tkatchenko, Alexandre},
title={Towards exact molecular dynamics simulations with machine-learned force fields},
journal={Nature Communications},
year={2018},
month={Sep},
day={24},
volume={9},
number={1},
pages={3887},
issn={2041-1723},
doi={10.1038/s41467-018-06169-2},
url={https://doi.org/10.1038/s41467-018-06169-2}
}

@article{sauceda20,
    author = {Sauceda, Huziel E. and Gastegger, Michael and Chmiela, Stefan and Müller, Klaus-Robert and Tkatchenko, Alexandre},
    title = "{Molecular force fields with gradient-domain machine learning (GDML): Comparison and synergies with classical force fields}",
    journal = {The Journal of Chemical Physics},
    volume = {153},
    number = {12},
    pages = {124109},
    year = {2020},
    month = {09},
    issn = {0021-9606},
    doi = {10.1063/5.0023005},
    url = {https://doi.org/10.1063/5.0023005},
}

@Article{unke21,
author={Unke, Oliver T.
and Chmiela, Stefan
and Sauceda, Huziel E.
and Gastegger, Michael
and Poltavsky, Igor
and Sch{\"u}tt, Kristof T.
and Tkatchenko, Alexandre
and M{\"u}ller, Klaus-Robert},
title={Machine Learning Force Fields},
journal={Chemical Reviews},
year={2021},
month={Aug},
day={25},
publisher={American Chemical Society},
volume={121},
number={16},
pages={10142-10186},
issn={0009-2665},
doi={10.1021/acs.chemrev.0c01111},
url={https://doi.org/10.1021/acs.chemrev.0c01111}
}

@misc{zhang20,
      title={Machine learning dynamics of phase separation in correlated electron magnets}, 
      author={Puhan Zhang and Preetha Saha and Gia-Wei Chern},
      year={2020},
      eprint={2006.04205},
      archivePrefix={arXiv},
      primaryClass={cond-mat.str-el}
}

@article{zhang21,
  title = {Arrested Phase Separation in Double-Exchange Models: Large-Scale Simulation Enabled by Machine Learning},
  author = {Zhang, Puhan and Chern, Gia-Wei},
  journal = {Phys. Rev. Lett.},
  volume = {127},
  issue = {14},
  pages = {146401},
  numpages = {7},
  year = {2021},
  month = {Sep},
  publisher = {American Physical Society},
  doi = {10.1103/PhysRevLett.127.146401},
  url = {https://link.aps.org/doi/10.1103/PhysRevLett.127.146401}
}

@Article{zhang23,
author={Zhang, Puhan and Chern, Gia-Wei},
title={Machine learning nonequilibrium electron forces for spin dynamics of itinerant magnets},
journal={npj Computational Materials},
year={2023},
month={Mar},
day={03},
volume={9},
number={1},
pages={32},
issn={2057-3960},
doi={10.1038/s41524-023-00990-0},
url={https://doi.org/10.1038/s41524-023-00990-0}
}

@article{zhang22b,
author = {Sheng Zhang  and Puhan Zhang  and Gia-Wei Chern },
title = {Anomalous phase separation in a correlated electron system: Machine-learning enabled large-scale kinetic Monte Carlo simulations},
journal = {Proceedings of the National Academy of Sciences},
volume = {119},
number = {18},
pages = {e2119957119},
year = {2022},
doi = {10.1073/pnas.2119957119},
URL = {https://www.pnas.org/doi/abs/10.1073/pnas.2119957119}}

@article{cheng23a,
  title = {Machine learning for phase ordering dynamics of charge density waves},
  author = {Cheng, Chen and Zhang, Sheng and Chern, Gia-Wei},
  journal = {Phys. Rev. B},
  volume = {108},
  issue = {1},
  pages = {014301},
  numpages = {15},
  year = {2023},
  month = {Jul},
  publisher = {American Physical Society},
  doi = {10.1103/PhysRevB.108.014301},
  url = {https://link.aps.org/doi/10.1103/PhysRevB.108.014301}
}

@article{cheng23b,
  title = {Convolutional neural networks for large-scale dynamical modeling of itinerant magnets},
  author = {Cheng, Xinlun and Zhang, Sheng and Nguyen, Phong C. H. and Azarfar, Shahab and Chern, Gia-Wei and Baek, Stephen S.},
  journal = {Phys. Rev. Res.},
  volume = {5},
  issue = {3},
  pages = {033188},
  numpages = {13},
  year = {2023},
  month = {Sep},
  publisher = {American Physical Society},
  doi = {10.1103/PhysRevResearch.5.033188},
  url = {https://link.aps.org/doi/10.1103/PhysRevResearch.5.033188}
}

@article{Ghosh24,
  title = {Kinetics of orbital ordering in cooperative Jahn-Teller models: Machine-learning enabled large-scale simulations},
  author = {Ghosh, Supriyo and Zhang, Sheng and Cheng, Chen and Chern, Gia-Wei},
  journal = {Phys. Rev. Mater.},
  volume = {8},
  issue = {12},
  pages = {123602},
  numpages = {16},
  year = {2024},
  month = {Dec},
  publisher = {American Physical Society},
  doi = {10.1103/PhysRevMaterials.8.123602},
  url = {https://link.aps.org/doi/10.1103/PhysRevMaterials.8.123602}
}

@article{Fan24,
  title = {Coarsening of chiral domains in itinerant electron magnets: A machine learning force-field approach},
  author = {Fan, Yunhao and Zhang, Sheng and Chern, Gia-Wei},
  journal = {Phys. Rev. B},
  volume = {110},
  issue = {24},
  pages = {245105},
  numpages = {18},
  year = {2024},
  month = {Dec},
  publisher = {American Physical Society},
  doi = {10.1103/PhysRevB.110.245105},
  url = {https://link.aps.org/doi/10.1103/PhysRevB.110.245105}
}

@article{tyberg25,
  title = {Machine learning force field model for kinetic Monte Carlo simulations of itinerant Ising magnets},
  author = {Tyberg, Alexa and Fan, Yunhao and Chern, Gia-Wei},
  journal = {Phys. Rev. B},
  volume = {111},
  issue = {23},
  pages = {235132},
  numpages = {11},
  year = {2025},
  month = {Jun},
  publisher = {American Physical Society},
  doi = {10.1103/d3zm-pbr1},
  url = {https://link.aps.org/doi/10.1103/d3zm-pbr1}
}

@misc{Jang25,
      title={Kinetics of Peierls dimerization transition: Machine learning force-field approach}, 
      author={Ho Jang and Yang Yang and Gia-Wei Chern},
      year={2025},
      eprint={2510.20659},
      archivePrefix={arXiv},
      primaryClass={cond-mat.stat-mech},
      url={https://arxiv.org/abs/2510.20659}, 
}

@article{Ma19,
  title = {Machine learning electron correlation in a disordered medium},
  author = {Ma, Jianhua and Zhang, Puhan and Tan, Yaohua and Ghosh, Avik W. and Chern, Gia-Wei},
  journal = {Phys. Rev. B},
  volume = {99},
  issue = {8},
  pages = {085118},
  numpages = {6},
  year = {2019},
  month = {Feb},
  publisher = {American Physical Society},
  doi = {10.1103/PhysRevB.99.085118},
  url = {https://link.aps.org/doi/10.1103/PhysRevB.99.085118}
}

@article{Liu22,
  title = {Machine learning predictions for local electronic properties of disordered correlated electron systems},
  author = {Liu, Yi-Hsuan and Zhang, Sheng and Zhang, Puhan and Lee, Ting-Kuo and Chern, Gia-Wei},
  journal = {Phys. Rev. B},
  volume = {106},
  issue = {3},
  pages = {035131},
  numpages = {12},
  year = {2022},
  month = {Jul},
  publisher = {American Physical Society},
  doi = {10.1103/PhysRevB.106.035131},
  url = {https://link.aps.org/doi/10.1103/PhysRevB.106.035131}
}

@article{Holstein1959,
    title={Studies of polaron motion: Part $\mathrm{\uppercase\expandafter{\romannumeral1}}$. The molecular-crystal model},
    author={Holstein, Th},
    journal={Annals of Physics},
    volume={8},
    number={3},
    pages={325--342},
    year={1959},
    publisher={Elsevier},
    doi={https://doi.org/10.1016/0003-4916(59)90002-8},
    url={https://www.sciencedirect.com/science/article/pii/0003491659900028}
}

@misc{Kingma2017,
      title={Adam: A Method for Stochastic Optimization}, 
      author={Diederik P. Kingma and Jimmy Ba},
      year={2017},
      eprint={1412.6980},
      archivePrefix={arXiv},
      primaryClass={cs.LG},
      url={https://arxiv.org/abs/1412.6980}, 
}

@article{Paszke2019,
  title={Pytorch: An imperative style, high-performance deep learning library},
  author={Paszke, Adam and Gross, Sam and Massa, Francisco and Lerer, Adam and Bradbury, James and Chanan, Gregory and Killeen, Trevor and Lin, Zeming and Gimelshein, Natalia and Antiga, Luca and others},
  journal={Advances in neural information processing systems},
  volume={32},
  year={2019},
  doi = {10.48550/arXiv.1912.01703},
  url = {https://arxiv.org/abs/1912.01703}
}

@article{Weber2018,
    title = {Two-dimensional Holstein-Hubbard model: Critical temperature, Ising universality, and bipolaron liquid},
    author = {Weber, Manuel and Hohenadler, Martin},
    journal = {Phys. Rev. B},
    volume = {98},
    issue = {8},
    pages = {085405},
    numpages = {8},
    year = {2018},
    month = {Aug},
    publisher = {American Physical Society},
    doi = {10.1103/PhysRevB.98.085405},
    url = {https://link.aps.org/doi/10.1103/PhysRevB.98.085405}
}

@inproceedings{Nair2010, 
author = {Nair, Vinod and Hinton, Geoffrey E.}, title = {Rectified Linear Units Improve Restricted Boltzmann Machines}, year = {2010}, isbn = {9781605589077}, publisher = {Omnipress}, address = {Madison, WI, USA}, booktitle = {Proceedings of the 27th International Conference on International Conference on Machine Learning}, pages = {807–814}, numpages = {8}, location = {Haifa, Israel}, series = {ICML'10} }
\end{document}